\documentclass[11pt]{article}

\usepackage[utf8]{inputenc}
\usepackage[T1]{fontenc}
\usepackage{lmodern}
\usepackage{orcidlink}
\usepackage{amsmath}
\usepackage{tabularx}
\usepackage{array}
\usepackage{graphicx}
\usepackage{placeins}
\usepackage{booktabs}
\usepackage{natbib}
\usepackage{url}
\usepackage{hyperref}
\usepackage{xurl}
\usepackage{fvextra}
\usepackage{adjustbox}

\usepackage{setspace}
\DeclareRobustCommand{\pkg}[1]{\texttt{#1}}
\DeclareRobustCommand{\proglang}[1]{\textsf{#1}}
\DeclareRobustCommand{\code}[1]{\texttt{\detokenize{#1}}}
\DeclareRobustCommand{\email}[1]{\href{mailto:#1}{\nolinkurl{#1}}}

\usepackage{fvextra}

\usepackage{fancyvrb}

\newenvironment{CodeChunk}{}{}

\DefineVerbatimEnvironment{CodeInput}{Verbatim}{
  fontsize=\scriptsize,
  baselinestretch=1.15
}

\DefineVerbatimEnvironment{CodeOutput}{Verbatim}{
  fontsize=\scriptsize,
  baselinestretch=1.15
}

\usepackage[a4paper,margin=2.8cm]{geometry}

\let\oldincludegraphics\includegraphics
\renewcommand{\includegraphics}[2][]{%
  \oldincludegraphics[width=\linewidth,keepaspectratio,#1]{#2}%
}

\newcommand{\Plainauthor}[1]{}
\newcommand{\Plaintitle}[1]{}
\newcommand{\Shorttitle}[1]{}
\newcommand{\Plainkeywords}[1]{}
\newcommand{\Address}[1]{}

\long\def\Abstract#1{\gdef\theabstract{#1}}
\long\def\Keywords#1{\gdef\thekeywords{#1}}

\providecommand{\tightlist}{%
  \setlength{\itemsep}{0pt}\setlength{\parskip}{0pt}}

\title{\pkg{hyreg2}: An \proglang{R} package to Estimate Latent Classes
on a Mixture of Continuous and Dichotomous Data}

\author{
Svenja Elkenkamp~\orcidlink{0000-0001-8704-1208}\\
Bielefeld University
\and
John Grosser~\orcidlink{0000-0003-3890-5596}\\
Bielefeld University
\and
Kim Rand~\orcidlink{0000-0001-7692-4099}\\
Akershus University Hospital
}

\date{}

\Plainauthor{Svenja Elkenkamp, John Grosser, Kim Rand}
\Plaintitle{hyreg2: An R Package to Estimate Latent Classes for a
Mixture of Continuous and Dichotomous Data}
\Shorttitle{\pkg{hyreg2}}

\Abstract{
The \proglang{R} package \pkg{hyreg2} introduces a frequentist framework
for estimating latent class models for mixed outcome types using a joint
likelihood approach. The method combines continuous and dichotomous data
under the assumption that both outcome types arise from a common
underlying data-generating process. In the implemented model, continuous
responses are assumed to follow a normal distribution, while dichotomous
responses are modeled using a binomial distribution. Such models are
used in various scientific disciplines to estimate a common set of
parameters across different types of data (e.g.~clinical trials,
econometrics and health economics). Latent class estimation is performed
using the expectation--maximization algorithm as implemented in the
widely used R package \pkg{flexmix}. The \pkg{hyreg2} package offers a
user-friendly implementation of this joint likelihood framework,
allowing users to estimate models without explicitly programming the
likelihood function. Heteroskedasticity as well as censored data can be
taken into account. In addition to model estimation, the package
provides dedicated summary and visualization functions to facilitate the
interpretation of results. The article presents the methodological
framework underlying the package and illustrates its functionality
through an example based on the estimation of an EQ-5D-5L value set.
}


\Plainkeywords{joint likelihood, latent class analysis, hybrid
model, continuous data, dichotomous data, R}

\Address{
Svenja Elkenkamp\\
Bielefeld University\\
Department of Health Economics and Health Care Management\\
Germany\\
E-mail: \email{svenja.elkenkamp@uni-bielefeld.de}\\

John Grosser\\
Bielefeld University\\
Department of Health Economics and Health Care Management\\
Germany\\

Kim Rand\\
Akershus University Hospital\\
Health Services Research Unit\\
Norway;\\
Maths in Health\\
The Netherlands
}

\begin{document}

\maketitle

\begin{abstract}
\theabstract
\end{abstract}


\bigskip

\section{Introduction}\label{introduction}

Statistical models that combine different types of data within a joint
likelihood - also known as hybrid models - are widely used across
disciplines such as marketing research, environmental studies, economics
and health sciences. Joint Likelihood models are in general particularly
useful in empirical settings where data arise from different measurement
scales but are assumed to be generated by the same underlying data
generation processes. By jointly modelling mixed outcome types, hybrid
approaches allow for more efficient estimation and a coherent
interpretation of model parameters compared to separate analyses of each
type of data \citep{Muthen.1984, Skrondal.2004}. One Example for joint
likelihood models in environmental and transport economics are Hybrid
Choice Models \citep{Benakiva.2002} , which can be used to combine
discrete choice data with psychological variables (e.g., environmental
attitudes) to explain behavioral decisions. In particular, a combination
of continuous and dichotomous data is of interest in many
decision-making contexts. For example, in studies of consumer behavior,
one may jointly model a binary decision to buy a product together with
an continuous willingness-to-pay value. In clinical research, various
studies jointly modeled these two types of data, such as continuous
efficacy measures and binary toxicity outcomes for medications
\citep{Aout.2018, Ezzalfani.2019}. Joint models for longitudinal binary
and continuous processes have also been used for analyzing smoking
cessation status and weight change simultaneously (e.g.
\citet{Liu.2009}).

One specific model, which combines continuous data with dichotomous
outcomes within a joint likelihood framework, was proposed by
\citet{RamosGoni.2017}. The model is estimated by maximizing this joint
likelihood, allowing both outcome types to contribute simultaneously to
inference. This model has been applied in numerous studies across
different countries to generate country-specific value sets of the
EQ-5D-5L questionnaire for use in health economic evaluations
\citep{Devlin.2022, Ramosgoni.2018, Ludwig.2018, Lin.2018, Rowen.2022}.
Although the model is originally motivated by estimating these value
sets, the underlying model structure is generic and applicable to a wide
range of problems involving jointly observed continuous and dichotomous
data.

In practice, Bayesian approaches are frequently employed to model joint
likelihoods, as they offer a flexible framework for integrating multiple
data sources and complex model structures. However, especially in health
economics and clinical trial settings, it can be highly challenging to
specify meaningful prior distributions when a method or intervention is
being evaluated for the first time, either in general or within a
specific country or contextual setting. In such situations, it is
particularly valuable to complement Bayesian approaches with frequentist
modeling strategies, as frequentist inference allows conclusions to be
drawn directly from the observed data without relying on potentially
speculative prior assumptions\citep{Casella.2002}. Additionally, it is
important that the modeling framework is also capable of identifying
latent classes within the data. Identifying latent classes can be
crucial, as failure to account for unobserved heterogeneity may lead to
biased estimates.Latent class models provide a flexible and widely used
approach to account for unobserved heterogeneity by assuming that the
population consists of a finite number of latent subgroups, each
characterized by distinct model parameters
\citep{Lazarsfeld.1968, Mclachlan.2000}.

Several software frameworks support the estimation of joint likelihood
models for outcomes following different distributions, including
combinations of continuous and dichotomous data. Bayesian frameworks
provide flexible tools for such models. In particular,
\pkg{brms}\citep{brms} and \pkg{MCMCglmm}\citep{mcmcglmm} allow the
joint modeling of continuous and dichotomous data within a multivariate
specification. These approaches construct the joint likelihood
internally but rely on Bayesian estimation. In contrast, non-Bayesian,
maximum-likelihood--based implementations are more limited. Generalized
linear mixed modeling tools such as \pkg{lme4}\citep{lme4} allow
different distributions but are restricted to single-outcome likelihoods
and therefore do not provide a joint likelihood across different types
of data. There are fully flexible maximum likelihood frameworks such as
Template Model Builder \pkg{TMB} \citep{tmb}, which permit arbitrary
combinations of likelihoods but they require users to explicitly program
the joint likelihood function. Hence, only advanced user are able to
implement a model as described by \citet{RamosGoni.2017} The
\proglang{R}\citep{RCoreTeam.2022} package \pkg{xreg} \citep{Rand.2016}
and the \proglang{Stata} package \pkg{hyreg} \citep{RamosGoni.2016}
implement the model proposed by \citet{RamosGoni.2017}; however, they do
not allow for the estimation of latent class models. Latent class models
in general are available in several \proglang{R} packages. The package
\pkg{lcmm} \citep{lcmm} focuses on latent class regression and mixed
models for longitudinal data, primarily with continuous or ordinal data,
while \pkg{poLCA} \citep{polca} provides latent class regression for
categorical responses only. The package \pkg{flexmix}
\citep{Leisch.2004, Grun.2008} offers a general
expectation-maximization(EM)-based framework for mixtures of regression
models and allows linear and logistic regressions. It is possible to
implement arbitrary distributions in the maximization-step (M-step) of
the EM-algorithm. In principle, this allows for the use of joint
likelihoods. However, no default implementation is provided, and
estimating joint likelihood models with latent classes in \pkg{flexmix}
requires substantial programming effort and advanced technical
expertise.

Hence, there is no existing software which combines joint likelihood
estimation based on continuous and dichotomous data with latent class
analysis on a user friendly level also for non experts in programming.
To address this gap, we introduce the new \proglang{R} package called
\pkg{hyreg2} following the framework proposed by \citet{RamosGoni.2017}.
The key methodological contribution of \pkg{hyreg2} lies in its
extension of joint likelihoods to a latent class framework using maximum
likelihood estimation. The package builds on the flexibility of
\pkg{flexmix} by using user-defined functions in the M-step as described
above. However, it remains user-friendly, as users of \pkg{hyreg2} are
not required to write likelihood functions themselves or to implement
them directly within the \pkg{flexmix} framework. In
addition,\pkg{hyreg2} provides further functionality for visualization
and summarization of results. The package also supports the analysis of
censored data, allows for heteroskedastic error structures, and enables
the estimation of partial coefficients as well as non-linear regression
models. By offering an accessible and extensible implementation in
\proglang{R}, the package lowers the barrier to applying latent class
models using joint likelihoods during maximum likelihood estimation.

In this article, chapter \ref{methods} introduces the underlying
methodology, describing the implemented joint likelihood model and the
latent class analysis. Chapter \ref{the-hyreg2-package} provides a
general overview of the structure and functionalities of the package and
outlines potential use cases. Subsequently, chapter
\ref{case-study-eq-5d-5l-value-set-estimation} illustrates the practical
use of the package through a case study on the estimation of an EQ-5D-5L
value set. Finally, chapter \ref{conclusion-and-outlook} explains
potential directions for further development of the package.

\section{Methods}\label{methods}

\subsection{Joint likelihood for continuous and dichotomous
data}\label{joint-likelihood-for-continuous-and-dichotomous-data}

\citet{RamosGoni.2017} describe a hybrid model that simultaneously
incorporates data following different distributions in order to maximize
a joint likelihood function. Specifically, the model combines continuous
data assumed to follow a normal distribution with dichotomous data
assumed to follow a binomial distribution. Equation \ref{eq:1} and
equation \ref{eq:2} show the likelihood formulas for normal and binomial
distributions in general. Thereby, \(y_i\) denotes the dependent
variable for observation \(i\) while \(X_i\) represents a vector
containing all independent variables for observation \(i\). The model
parameter vector \(\beta\) is associated with the continuous (\(C\)) and
\(\beta^\ast\) with the dichotomous (\(D\)) data.

\begin{equation}
\begin{aligned}
L_{\text{normal}}(\beta,\sigma)
&= \prod_{y_i \in C} f_{\text{normal}}(y_i \mid X_i, \beta, \sigma) \\
&= \prod_{y_i \in C}
\frac{1}{\sqrt{2\pi\sigma^2}}
\exp\left(
-\frac{(y_i - X_i\beta)^2}{2\sigma^2}
\right)
\end{aligned}
\label{eq:1}
\end{equation}

\vspace{1.5em}

\begin{equation}
\begin{aligned}
L_{\text{binomial}}(\beta^\ast)
&= \prod_{y_i \in D} f_{\text{binomial}}(y_i| X_i, \beta^\ast) \\
&=
\prod_{y_i \in \textit{D}}
\left(\frac{1}{1 + e^{-X_i\beta^\ast}}\right)^{y_i}
\left(\frac{e^{-X_i\beta^\ast}}{1 + e^{-X_i\beta^\ast}}\right)^{1 - y_i}
\end{aligned}
\label{eq:2}
\end{equation}

The continuous and dichotomous data are linked through a shared set of
model parameters and estimated jointly under the assumption of
conditional independence. Therefore, the joint likelihood is given by
the product of the normal and binomial likelihoods and maximum
likelihood estimation is performed by maximizing this product with
respect to the model parameters \citep{Casella.2002}. For
interpretational purposes, it can be useful to rescale the estimated
parameter vector \(\beta\) to the scale of the binomially distributed
parameters \(\beta^\ast\). Accordingly, an additional scaling parameter
\(\theta\) is introduced, for which holds
\(\beta = \theta * \beta^\ast\).

\begin{equation}
\begin{aligned}
L( \beta,\sigma, \theta)
&=
L_{\text{normal}}(\beta, \sigma) * L_{\text{binomial}}( \beta^\ast) \\
&= L_{\text{normal}}(\beta, \sigma) * L_{\text{binomial}}(\frac{\beta}{\theta} )
\end{aligned}
\label{eq:3}
\end{equation}

However, in practice, estimation is carried out by maximizing the sum of
the log-likelihood functions rather than the product of the individual
likelihoods. Hence, the log likelihood function to be used in
\pkg{hyreg2} is given in equation \ref{eq:4}.

\begin{equation}
\begin{aligned}
\log L(\beta,\sigma,\theta)
&=
-\tfrac{1}{2} \sum_{y_i \in C}
\left[ \log(2\pi\sigma^2) + \frac{(y_i - X_i\beta)^2}{\sigma^2} \right] \\
&\quad + \sum_{y_i \in D}
\left[
y_i \log\left(\frac{1}{1 + e^{-X_i \frac{\beta}{\theta}}}\right)
+ (1 - y_i) \log\left(\frac{e^{-X_i \frac{\beta}{\theta}}}{1 + e^{-X_i\frac{\beta}{\theta}}}\right)
\right]
\end{aligned}
\label{eq:4}
\end{equation}

\(X_i\) is the vector containing all independent variable observations
of row \(i\), and \(\beta\) is the column vector of the joint model
parameter estimates. Additionally the parameters \(\sigma\) and
\(\theta\) have to be estimated. More detailed information about this
log likelihood formula and its extension for censored data can be found
in \citet{RamosGoni.2017}.

\subsection{Latent Class Analysis and
Expectation-Maximization-Algorithm}\label{latent-class-analysis-and-expectation-maximization-algorithm}

Latent class analysis (LCA) is based on the assumption that unobserved
membership of groups can be determined by certain patterns in the data
\citep{Weller.2020}. In addition to group membership, other parameters
that differ between the groups can be estimated using latent class
regression \citep{Schreiber.2022}. One possibility for estimating latent
class models is using the EM-algorithm first described by
\citet{Dempster.1977}. This algorithm is a repeated procedure, which
consists of an expectation-step (E) and a maximization-step (M).
Different descriptions and modifications of the EM-algorithm can be
found in the literature \citep{Fox.2016}. In general, the posterior
class probabilities for each observation are estimated during the
E-step, while the log-likelihood is computed for each class separately
using these probabilities during the M-step \citep{Leisch.2004}. The
estimated parameters from the M-step are then used to adjust the class
probabilities in a new E-step and the process is repeated until
convergence of the parameters for each class is reached. Since the
latent class estimation of \pkg{hyreg2} is based on \pkg{flexmix} source
code, the following description of LCA and the EM-algorithm are based on
the descriptions by the authors of \pkg{flexmix}, as given in
\citet{Leisch.2004}.

Latent class regression models can generally be described as a mixture
(or combination) of standard linear regression models. Consider a data
set with \(N\) observations and \(K\) latent classes.In \pkg{hyreg2}, we
want to estimate a finite mixture model of the form

\begin{equation}
h(y_i\mid X_i,\psi)=\sum_{k=1}^{K}\pi_k f(y_i \mid X_i,\beta_k, \sigma_k, \theta_k)
\label{eq:4a}
\end{equation}

where \(y_i\) is the dependent variable with the conditional density
\(h\), \(X_i\) is a vector of independent variables with
\(i\in\left\{1,\ldots, N\right\}\). Furthermore, \(\pi_k\) is the prior
probability of class \(k\) and all \(\pi_k\) have to sum to one.
Moreover, \(\beta_k\) is the latent class specific parameter vector and
\(\sigma_k\) and \(\theta_k\) are the latent class specific \(\sigma\)
and \(\theta\) parameters from equation \ref{eq:3} and \ref{eq:4} for
the density function \(f\) of class \(k\).

\begin{equation}
\psi = (\pi_1,\ldots,\pi_K,\beta_1,\ldots,\beta_K, \sigma_1,\ldots,\sigma_K, \theta_1,\ldots,\theta_K)
\end{equation}

Therefore, \(\psi\) is the vector of all parameters to be estimated with
\(k\in\left\{1,\ldots, K\right\}\).

The log-likelihood of a sample of \(N\) observations
\(\{(X_1, y_1),\ldots,(X_N, y_N)\}\) is in general given by

\begin{equation}
\begin{aligned}
\log L
&= \sum_{i=1}^{N}\log\left(h(y_i \mid X_i,\psi)\right) \\
&= \sum_{i=1}^{N}\log\left(\sum_{k=1}^{K}\pi_k f(y_i \mid X_i,\beta_k, \sigma_k, \theta_k)\right)
\end{aligned}
\label{eq:5}
\end{equation}

and usually cannot be maximized directly. The vector \(\psi\),
containing the prior class probabilities \(\pi_k\), the parameters
\(\sigma_k\) and \(\theta_k\) and the parameter vectors \(\beta_k\),
\(k\in\left\{1,\ldots, K\right\}\), can then be estimated using the
EM-algorithm.

First, one provides a set of starting values for the prior class
probabilities \(\pi_k\), the parameters \(\sigma_k\) and \(\theta_k\)
and the parameter vectors \(\beta_k\),
\(k\in\left\{1,\ldots, K\right\}\). Then the algorithm consists of two
steps, which are repeated iteratively until convergence is reached: the
E-step and the M-step.

\vspace{1.5em}

\textbf{E-step}

Calculate the conditional probability that the \(i\)-th data point
belongs to one specific class \(k\), where \(i\in\{1,\ldots, N\}\),
\(k\in\{1,\ldots, K\}\)

\begin{equation}
P(X_i \text{ belongs to class } k \mid X_i, y_i, \psi)
=
\frac{\pi_k f(y_i \mid X_i,\beta_k, \sigma_k, \theta_k)}{\sum_{j=1}^{K}\pi_j f(y_i \mid X_i,\beta_j, \sigma_j, \theta_j)}
= p_{ik}
\label{eq:6}
\end{equation}

Using equation \ref{eq:6} for each observation \(i\) and each class
\(k\), the class probabilities for each class \(k\) can be updated.

\begin{equation}
\pi_k=\frac{1}{N}\sum_{i=1}^{N} p_{ik}
\label{eq:7}
\end{equation}

\vspace{1.5em}

\textbf{M-step}

Maximize the log-likelihood for each class separately using the
posterior probabilities \(\hat{p}_{ik}\) from the E-step as weights.

\begin{equation}
\max_{\beta_k} \sum_{i=1}^{N} p_{ik}\log f(y_i|X_i,\beta_k, \sigma_k, \theta_k)
\label{eq:8}
\end{equation}

The \(\sigma_k\), \(\theta_k\) and \(\beta_k\) which yield this
maximization are then used in a new iteration of the E-step.

In this manner, the E-step and M-step are repeated until the likelihood
improvement falls under a pre-specified threshold or a maximum number of
iterations is reached. Although the EM-algorithm is a good way to
predict latent classes in the data, it also has limitations. First, it
can take a long time to reach convergence, requiring significant
computational power. Second, as with all maximum likelihood estimations,
it is possible that local (rather than global) maxima are reached,
causing the algorithm to stop prematurely. This may depend on the chosen
starting values. In such cases, parameter values may be found that are
not the true parameters. Therefore, it is advisable to run the algorithm
multiple times, preferably with different starting values, and compare
the results. Third, problems in parameter estimation arise if a
component is assigned only a few observations.

\section{The hyreg2 package}\label{the-hyreg2-package}

\subsection{Overview}\label{overview}

The R package \pkg{hyreg2} provides an extensible framework for
estimating latent classes using a joint likelihood model that combines
continuous and dichotomous data within a unified likelihood. The package
is available on The Comprehensive \proglang{R} Archive Network (CRAN)
\citep{cran}. It is built using the latent class estimation of the
widely used \proglang{R} package \pkg{flexmix}, leveraging its EM-based
estimation framework. The joint likelihood is implemented via a custom
M-step driver, which is seamlessly integrated into the \pkg{flexmix}
workflow. Therefore, users of \pkg{hyreg2} do not need to possess
in-depth programming skills. Parameter optimization is automativcally
carried out using maximum likelihood estimation via the \texttt{mle2}
function from the \pkg{bbmle} package \citep{bbmle}, allowing for
flexible choice of optimizers and convergence controls.

The package provides two main user-facing functions: \texttt{hyreg2} and
\texttt{hyreg2\_het}. While \texttt{hyreg2} estimates models under the
assumption of homoscedastic errors in the continuous data part,
\texttt{hyreg2\_het} extends this framework by allowing for
heteroscedasticity through a separate variance model for the continuous
data. Both functions support censored data, class-specific starting
values, and flexible specification of covariates across the continuous
and dichotomous data. In addition, non-linear model formulas can be
specified for homoscedastic models. This can be achieved by defining the
model formula as a combination of parameters and variables, as also used
by e.g.~the \texttt{nls} function from the \pkg{stats} package
\citep{rstats} and described in chapter
\ref{using-non-classic-model-formulas} of this article. Additional
utility functions, including \texttt{summary\_hyreg2},
\texttt{plot\_hyreg2}, \texttt{give\_id}, and \texttt{get\_stv},
facilitate model inspection, visualization of class assignments, and
reuse of parameter estimates as starting values in subsequent model
runs.

If the number of classes to be estimated is set to one, \pkg{hyreg2}
produces results equivalent to what can be achieved using existing
software which also implement the model by \citet{RamosGoni.2017},
e.g.~\pkg{xreg} (\proglang{R}) or \pkg{hyreg} (\proglang{STATA}).

\begin{figure}[ht]
\centering
\includegraphics[width=0.9\linewidth]{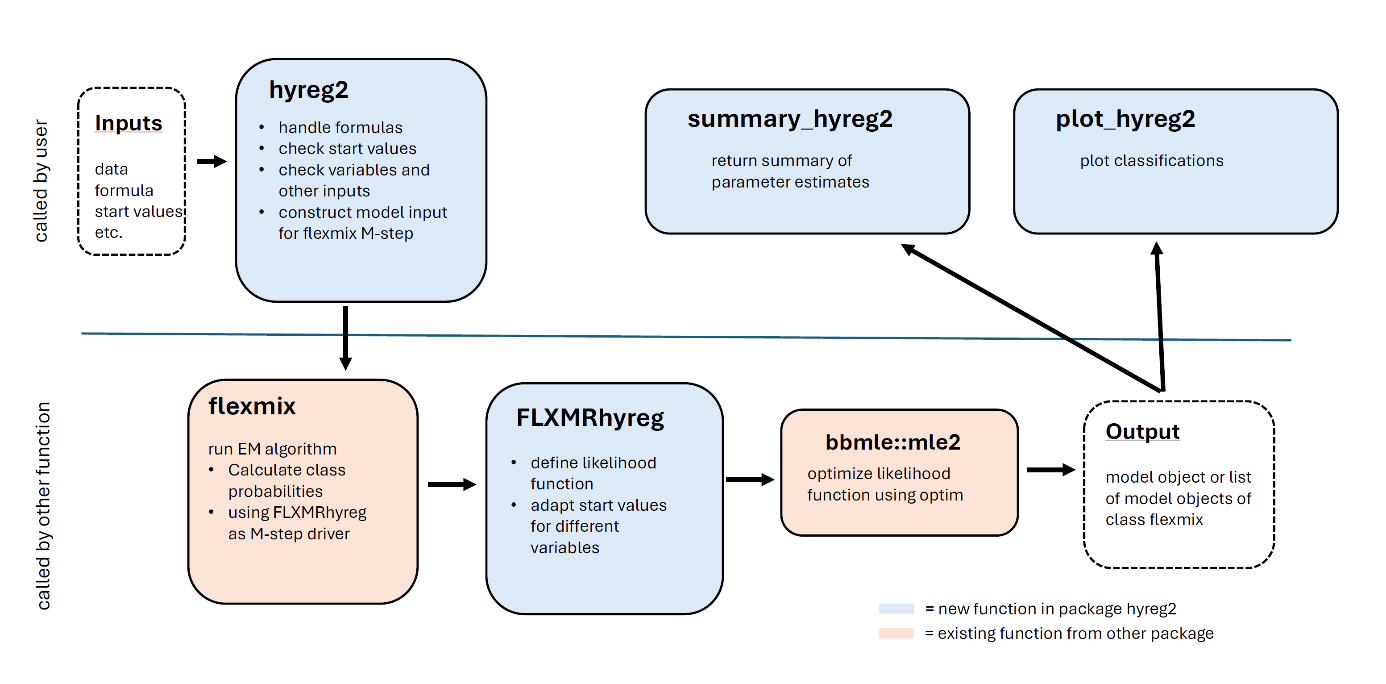}
\caption{structure of the hyreg2 package}
\label{fig1}

\end{figure}

Figure \ref{fig1} visualizes the model estimation process for the
\texttt{hyreg2}function. The process begins when the user calls
\texttt{hyreg2} and specifies the required input variables. The function
first validates the inputs, checking for misspecifications or missing
values. It then constructs a model object, which is passed to
\texttt{flexmix}. At this stage, \texttt{flexmix} is executed, running
the EM-algorithm. The E-step is handled entirely by the original
\pkg{flexmix} source code. During the M-step, the custom driver
\texttt{FLXMRhyreg} is invoked. This driver defines the joint likelihood
and calls \texttt{bbmle:mle2} to optimize the log-likelihood and compute
the parameter estimates. The function returns an output object of type
\texttt{flexmix}, either a single \texttt{model} or a \texttt{list} of
models, depending on the input specification. This output can be passed
to \texttt{summary\_hyreg2} or \texttt{plot\_hyreg2} by the user for
further analysis and visualization.

\subsection{Functionality}\label{functionality}

At this point, the use and functionality of the functions
\texttt{hyreg2} and \texttt{hyreg2\_het} is introduced according to the
help page available in the package. More detailed information about how
they work and how user should specify the different input parameters are
demonstrated using a simulated data set in the package vignette
available on CRAN (\url{https://CRAN.R-project.org/package=hyreg2}).
Moreover, chapter \ref{case-study-eq-5d-5l-value-set-estimation} of this
article demonstrates the functionality using real-world data and
provides explicit code examples.

When using \texttt{hyreg2} or \texttt{hyreg2\_het} to estimate latent
class models, it is essential to account for groups of observations.
Controlling for such dependencies is essential to ensure valid parameter
estimation and inference. Otherwise, the model will put the majority of
continuous data into one class and the majority of dichotomous data into
the other. Hence, the model will not detect latent classes on the
mixture of the data but will separate the two different types of data.
This behavior can be replicated using the vignette.

\begin{table}[!htbp]
\centering
\begin{tabular}{p{3cm} p{12cm}}
\hline
\textbf{Parameter} & \textbf{Description} \\
\hline
\texttt{formula} & Model \texttt{formula}, can be linear or non-linear. For non-linear formulas, variables and parameters must be provided and \texttt{formula\_type\_classic} must be set to \texttt{FALSE}. Using \texttt{|xg} will include a grouping variable \texttt{xg}. \\
\texttt{data} & A data frame containing the data. \\
\texttt{type} & Either the name of the column in \texttt{data} containing an indicator of whether an observation is continuous or dichotomous, or a vector containing the indicator. \\
\texttt{type\_cont} & Value of \texttt{type} referring to continuous data. \\
\texttt{type\_dich} & Value of \texttt{type} referring to dichotomous data. \\
\texttt{k} & Number of latent classes to be estimated via \texttt{flexmix}. \\
\texttt{control} & Control list for \texttt{flexmix}. \\
\texttt{stv} & Start values for all coefficients in the formula, including sigma and theta. \\
\texttt{offset} & Offset as in \texttt{flexmix}. \\
\texttt{optimizer} & Optimizer to be used in \texttt{mle2}, default \texttt{"optim"}. \\
\texttt{opt\_method} & Optimization method to be used in optimizer, default \texttt{"BFGS"}. \\
\texttt{lower} & Lower bound for censored data, default \texttt{-Inf}. If used, \texttt{opt\_method} must be set to \texttt{"L-BFGS-B"}. \\
\texttt{upper} & Upper bound for censored data, default \texttt{Inf}. If used, \texttt{opt\_method} must be set to \texttt{"L-BFGS-B"}. \\
\texttt{latent} & Data type to use in component identification; must be one of \texttt{"both"}, \texttt{"cont"} or \texttt{"dich"}, default \texttt{"both"}. \\
\texttt{id\_col} & Name of the grouping variable; only needed if \texttt{latent != "both"}. \\
\texttt{classes\_only} & Indicates whether the function should perform only classification rather than both classification and model estimation; only possible for \texttt{latent != "both"}. \\
\texttt{variables\_both} & Variables to be fitted on both continuous and dichotomous data. If not specified, all variables from \texttt{formula} are used. If provided and not all variables from \texttt{formula} are included, \texttt{variables\_cont} and \texttt{variables\_dich} must be provided as well (one of them can be \texttt{NULL}). \\
\texttt{variables\_cont} & Variables to be fitted only on continuous data. If provided, \texttt{variables\_both} and \texttt{variables\_dich} must be provided as well. \\
\texttt{variables\_dich} & Variables to be fitted only on dichotomous data. If provided, \texttt{variables\_both} and \texttt{variables\_cont} must be provided as well. \\
\texttt{formula\_\allowbreak type\_\allowbreak classic} & Is the provided \texttt{formula} a classic R formula containing only variables (\texttt{TRUE}) or does it include both variables and parameters (\texttt{FALSE})? Default \texttt{TRUE}. \\
\texttt{...} & Additional arguments for \texttt{flexmix} or \texttt{bbmle2}. \\
\hline
\end{tabular}
\caption{hyreg2 arguments}
\label{tab:function-arguments}
\end{table}

Table \ref{tab:function-arguments} shows the arguments to be used in
\texttt{hygre2}. For some arguments further details are given below. As
an example data set the included data set \texttt{simulated\_data\_norm}
can be used to follow the explanations given below.

\vspace{1.5em}

\textbf{formula}

A classic R formula containing only variables
(e.g.~\texttt{y\ \textasciitilde{}\ x1\ +\ x2\ +\ ...}) can be provided
as well as a formula including variables and parameters (non-classic),
e.g.~\texttt{y\ \textasciitilde{}\ x1\ *\ beta1\ +\ x2\ *\ beta2} or
\texttt{y\ \textasciitilde{}\ 1/exp(x1\ *\ beta1\ +\ x2\ *\ beta2)},
where \texttt{beta} are the parameters to be estimated and the
\texttt{x}s are column names from the data set. In \pkg{hyreg2},
however, non-linear models can only be estimated using a non-classic
formula. If the provided formula is non-classic,
\texttt{formula\_type\_classic} must be set to \texttt{FALSE}. When
estimating an intercept, the formula must explicitly include a parameter
named \texttt{"INTERCEPT"} (without a corresponding variable from the
data set). Additionally, it is possible to include a grouping variable
for repeated measures by using \texttt{\textbar{}\ xg} where \texttt{xg}
is the column containing the group-memberships. The resulting formula
will look like this:
\texttt{y\ \textasciitilde{}\ x1\ +\ x2\ +\ ...\ \textbar{}\ xg}. In
\texttt{flexmix}, this is called the concomitant variable specification:
the model is fit conditional on grouping, so that all observations with
the same group are treated as belonging together when computing
likelihood contributions. One possible grouping variable can be an id
number to identify answers by the same participants. We highly recommend
using a grouping variable, since otherwise the algorithm for
\texttt{k\ =\ 2} tends to classify all continuous data into one
estimated class and all dichotomous data into the other.

\vspace{0.5em}

\textbf{data}

A dataframe having the following columns: all independent variables
(\texttt{x}) and the dependent variable \texttt{y} used in
\texttt{formula}, one column for the grouping variable \texttt{xg} if
grouping should be used (e.g.~ID numbers of participants with repeated
measurements), one column indicating if the observations belong to
continuous or dichotomous data with the entries \texttt{type\_cont} and
\texttt{type\_dich} (e.g., for a column called \texttt{"type"} with the
entries \texttt{"TTO"} for continuous data points and \texttt{"DCE"} for
dichotomous data points, \texttt{type\_cont} will be \texttt{"TTO"} and
\texttt{type\_dich} will be \texttt{"DCE"}). One row should match one
observation (one data point).

\vspace{1.5em}

\textbf{start values (stv)}

If the same start values \texttt{stv} are to be used for all latent
classes, the given start values must be a named vector. Otherwise (if
different start values are assumed for each latent class), a list of
named vectors should be used. In this case, there must be one entry in
the list for each latent class. Each start value vector must include
start values for sigma and theta. Currently, it is necessary to use the
names \texttt{"sigma"} and \texttt{"theta"} for these values. If users
are unsure for which variables start values must be provided (in the
linear formula case), this can be checked by calling
\code{colnames(model.matrix(formula, data))}. In this call, the
\texttt{formula} should not include the grouping variable.

\vspace{1.5em}

\textbf{latent, id\_col, classes\_only}

In some situations, it can be useful to identify the latent classes on
only one \texttt{type} of data while estimating the model parameters on
both \texttt{types} of data. In such cases, the input variable
\texttt{latent} can be used to specify on which type of data the
classification should be done. If \texttt{"cont"} or \texttt{"dich"} is
used, \texttt{formula} must contain a grouping variable and additionally
the input parameter \texttt{id\_col} must be specified and gives the
name of the grouping variable for classification. Some groups may be
removed from the data, since they have only continuous or only
dichotomous observations. Then in a first step, a model is estimated
only on the continuous/dichotomous data and the achieved classification
is stored. In a next step, model parameters are estimated separately for
each identified class on both \texttt{types} of data using this
classification. The output object of \texttt{hyreg2} in this case is a
list of k models. Additionally, at position k+1 of the list, a data
frame containing the corresponding classifications from the first step
is returned. Each element k in the list contains the estimated
parameters for one of the latent classes. When setting the input
variable \texttt{classes\_only} to \texttt{TRUE}, the second step is
left out and the estimated classes from step one are given as output.

\vspace{1.5em}

\textbf{variables\_both, variables\_cont, variables\_dich}

It is possible to specify partial coefficients, which are used only on
continuous or dichotomous data

\emph{Example:}

Suppose different models should be specified for continuous and
dichotomous data:

\begin{itemize}
\tightlist
\item
  Model continuous data: \texttt{y\ \textasciitilde{}\ x1\ +\ x3}
\item
  Model dichotomous data: \texttt{y\ \textasciitilde{}\ x1\ +\ x2}
\end{itemize}

The \texttt{formula} input to \texttt{hyreg2} must then include all
parameters that occur in either model:

\texttt{y\ \textasciitilde{}\ x1\ +\ x2\ +\ x3}

The assignment of parameters to data types is then achieved via the
input arguments \texttt{variables\_both}, \texttt{variables\_cont}, and
\texttt{variables\_dich}:

\begin{itemize}
\tightlist
\item
  \texttt{variables\_both\ =\ "x1"}
\item
  \texttt{variables\_cont\ =\ "x3"}
\item
  \texttt{variables\_dich\ =\ "x2"}
\end{itemize}

Every variable included in the provided \texttt{formula} (except the
grouping variable) must appear in exactly one of these vectors. One of
the \texttt{variables\_} vectors can also be \texttt{NULL}, if no
variables should be used only on this type of the data.

\vspace{1.5em}

\textbf{return object}

\texttt{hyreg2} will return a model object of type~\texttt{flexmix},
when using \texttt{latent\ =\ "both"} or~ a \texttt{list}~of model
objects of type~\texttt{flexmix} when using \texttt{latent\ =\ "cont"}
or \texttt{latent\ =\ "dich"}. Please note, that the estimates
for~\texttt{sigma}~and~\texttt{theta}~are on a log-scale and have to be
transformed using~\texttt{exp()}to get the correct estimated values.

\vspace{1.5em}

The \texttt{hygre2\_het} function allows controlling for
heterosecasticity and has the same arguments as the \texttt{hyreg2}
function except \texttt{formula\_type\_classic} since there is actually
no option to use non-classic formulas in \texttt{hyreg2\_het}. Using
\texttt{hyreg2\_het} additonal input parameters for the variance model
of the continuous data can be given , see table \ref{tab:parameterhet}.

\vspace{1.5em}

\begin{table}[htbp]
\centering
\begin{tabular}{p{3.5cm} p{11cm}}
\hline
\textbf{Parameter} & \textbf{Description} \\
\hline
\texttt{formula\_sigma} & Linear \texttt{formula} for sigma estimation. If \texttt{formula\_sigma} is not provided, \texttt{formula} (excluding any grouping variables) is used by default.\\

\texttt{stv\_sigma} & \texttt{named vector} with start values for sigma estimation. Names must correspond to the variables as given in \texttt{formula\_sigma}. \\
\hline
\end{tabular}
\caption{Additional hyreg2\_het arguments}
\label{tab:parameterhet}
\end{table}

\vspace{1.5em}

\textbf{formula\_sigma, stv\_sigma}

To account for heteroscedasticity in the data, an additional formula
\texttt{formula\_sigma} and an additional vector of starting values for
this formula (\texttt{stv\_sigma}) can be specified. The provided
\texttt{formula\_sigma} must be linear and the vector
\texttt{stv\_sigma} must contain start values for all parameters used in
the formula. If neither \texttt{formula\_sigma} nor \texttt{stv\_sigma}
are provided, the same inputs as for \texttt{formula} (without
controlling for groups) and \texttt{stv} (without sigma) are used. The
estimates for the \texttt{sigma} model can be identified in the model
output by the ending \texttt{"\_h"}. It is important to note that, when
using \texttt{hyreg2\_het}, neither \texttt{stv} nor \texttt{stv\_sigma}
are allowed to include a value called ``sigma'', because \texttt{sigma}
is estimated with its own formula (in contrast to \texttt{hyreg2}, where
``sigma'' must always be specified in \texttt{stv}).

\subsection{Use Cases}\label{use-cases}

\textbf{Econometrics and Marketing}

In many applications in marketing and consumer demand analysis,
purchasing behavior is understood as a multi-stage decision process.
Consumers first decide whether to purchase a product at all and
subsequently determine how much to purchase
\citep{Malhotra.2019, Solomon.2020}. These two aspects of behavior are
typically represented by different types of outcomes: a binary purchase
incidence indicator describing whether a purchase occurs, and a
continuous outcome describing the quantity or expenditure associated
with the purchase. Because both decisions are driven by the same
underlying preferences and covariates, modelling them separately may
ignore important information contained in the joint data structure. As a
result, econometric models have been developed that jointly analyse
discrete and continuous outcomes within a unified likelihood framework.
For example \citet{Manchanda.1999} analyse household purchasing behavior
across multiple product categories. In contexts such as consumer demand
modeling, where different groups of individuals may follow distinct
preference patterns or decision rules, latent class models provide a
flexible framework for identifying heterogeneous behavioral segments
while jointly exploiting information from multiple outcome types.
Consequently, this represents a meaningful application case for the
package \pkg{hyreg2}.

\vspace{1.5em}

\textbf{Clinical Trials}

In clinical trials, it is common that relevant study outcomes are
measured on different scales, such as dichotomous endpoints (e.g.,
occurrence of toxicity or treatment response) and continuous outcomes
(e.g., biomarkers or tumour size reduction). This is particularly
relevant in early-phase dose-escalation studies, where both safety and
efficacy need to be considered simultaneously. Accordingly, several
methodological contributions propose joint modelling approaches that
combine binary toxicity indicators with continuous efficacy measures in
order to support dose selection based on an integrated benefit--risk
assessment. Several studies have applied such approaches to
simultaneously analyse different types of data
\citep{Aout.2018, Ezzalfani.2019}. To date, these models have often been
implemented within a Bayesian framework. However, it remains unclear
whether this reflects a deliberate methodological preference for
Bayesian inference in all cases. An alternative explanation is that the
lack of accessible tools for frequentist implementation of such joint
models has contributed to the predominance of Bayesian approaches in the
literature. The \pkg{hyreg2} package addresses this gap by providing a
straightforward frequentist framework for estimating hybrid models,
thereby enabling even less experienced programmers to apply frequentist
joint modelling techniques. Moreover, \pkg{hyreg2} facilitates the
estimation of latent class joint likelihood models, which may be
particularly relevant in clinical applications. Without accounting for
latent groups in the data, subgroup-specific effects may be masked in
aggregated models, potentially leading to misleading conclusions
regarding both efficacy and safety.

\vspace{1.5em}

\textbf{Health Economics}

Quality-adjusted life years (QALYs) have become a cornerstone of health
economic evaluation, providing a standardized measure to assess the
value of healthcare interventions by integrating both the quantity and
quality of life. They are designed to capture changes in health-related
quality of life (HRQoL) resulting from medical treatments or public
health initiatives and are widely used in economic evaluations and
health technology assessments. QALYs are typically derived from health
state utility values, which reflect individuals' preferences for
different health states. One of the most widely used instruments for
measuring HRQoL is the EQ-5D-5L, developed by the EuroQol Group
\citep{Devlin.2022}. In particular, the EQ-5D-5L is the most commonly
applied instrument worldwide for informing QALY calculations
\citep{Wislff.2014}, owing to its robust methodology and widespread
adoption in clinical trials and health economic studies. To translate
EQ-5D health states into utility values, valuation studies often rely on
multiple elicitation methods, such as time trade-off (TTO) tasks and
discrete choice experiments (DCE). Therefore, the resulting data sets
consist of continuous and dichotomous data. Joint models as introduced
by \citet{RamosGoni.2017} play a central role in the construction of
value sets required for QALY computation. Within the EQ-5D community,
there has been growing interest in better accounting for preference
heterogeneity in the continuous data \citep{RamosGoni.2022}. This
interest is based on the assumption that participants may form more or
less distinct groups with similar preference profiles, which may be
better reflected using latent class models. Hence, the \pkg{hyreg2}
package provides a useful approach for estimating such models.

\section{Case study: EQ-5D-5L value set
estimation}\label{case-study-eq-5d-5l-value-set-estimation}

In this chapter, we illustrate some functionalities of the \pkg{hyreg2}
package using a detailed example. In particular, we use a publicly
available data set containing valuation study data for the EQ-5D-5L
questionnaire. We first introduce some background information about the
EQ-5D and the data set. Afterwards, we compare the results obtained with
\pkg{hyreg2} to those from the \proglang{R} package \pkg{xreg} when no
latent classes are estimated (\texttt{k\ =\ 1}). Since the R package
\pkg{xreg} also implements the model proposed by \citet{RamosGoni.2017},
the results are expected to coincide. Subsequently, we examine the
estimation of two latent classes and the visualization of results using
\code{hyreg2_plot}.Furthermore, we demonstrate how censored data or
heteroscedasticity can be taken into account. Finally, we demonstrate
the estimation of a non-classic model formula. More information about
other package functionalities (e.g., how start values can be provided as
a list or how partial coefficients can be estimated) can be found in the
package's vignette.

\vspace{1.5em}

The EQ-5D-5L captures five dimensions of health: mobility (mo),
self-care (sc), usual activities (ua), pain/discomfort (pd), and
anxiety/depression (ad)---each with five severity levels, with the
numbers 1-5 corresponding to no, slight, moderate, severe, or extreme
problems, respectively. The resulting descriptive system allows for the
classification of 3,125 unique health states \citep{Herdman.2011}.
Responses from the EQ-5D-5L questionnaire are converted into utility
values using country-specific value sets, which reflect the preferences
of the general population. The aim of such a value set is it to provide
a single utility value for each possible health state
\citep{Devlin.2022}. These utility values allow for the determination of
QALYs, which are needed for health economic evaluations
\citep{Drummond.2005}. The generated utility values cannot be higher
than 1 (full health), but utility values less than 0 (dead) are possible
and indicated that the related health states are, on average, seen as
worse than death \citep{Devlin.2022}.

To generate EQ-5D value sets, standardized valuation protocols have been
developed that specify how data should be collected
\citep{Oppe.2014, Stolk.2019}. The valuation studies usually consist of
TTO (continuous data) and DCE (dichotomous data) parts. Very briefly,
the TTO part involves asking the participant to choose between two
alternative hypothetical lives, one consisting of a fixed length of
time, e.g.~10 years, in an impaired state of health, and one consisting
of a shorter life span in full health. The length of life in full health
is varied until preferential indifference is reached, at which point the
value of the impaired state can be calculated. The resulting TTO values
range from -1 (trading all the lead time) to 1 (trading no years in full
health) \citep{Torrance.1972, Oppe.2016}, although these values are
usually linearly rescaled to the interval {[}0;2{]} for value set
generation. In contrast, DCE involves presenting participants with a
choice between two different health states, and modeling preferences by
way of conditional logistic regression \citep{Stolk.2010}. The chosen
health state is coded as 1 and the other as 0. Afterwards the resulting
continuous and dichotomous data are used to generate the country
specific value set using the joint likelihood framework by
\citet{RamosGoni.2017}. Therefore, a regression model must be estimated.

One of the most common ones is the so called 20-parameter model
\citep{Rowen.2022, Devlin.2022} as given in equation \ref{eq.eq5d}.

\begin{equation}
value = \beta_0 + mo2 * \beta_{mo2} + mo3 * \beta_{mo3} + \ldots + ad5 * \beta_{ad5} + \epsilon
\label{eq.eq5d}
\end{equation}

\vspace{1.5em}

In this case study, we use the publicly available EQ-5D data set
\citep{Rand.2018} to demonstrate the functionality of \texttt{hyreg2}
for the context of a full EQ-5D data set. This data set can be installed
to be used in \proglang{R} via GitHub using
\linebreak \texttt{devtools::install\_github("intelligentaccident/EQ5D\_data")}.
This data set combines valuation data from TTO and DCE tasks, following
the approach used in many EQ-5D valuation studies. Prior to model
estimation, the EQ-5D valuation data have to be cleaned and prepared
following the standard procedures described in the international
EQ-5D-5L valuation protocol \citep{Oppe.2014, Stolk.2019, Rowen.2022}.
Observations flagged during the quality-control process are excluded
(\texttt{fb\_flagged\ ==\ 0}) to remove responses that did not meet the
protocol's interview quality criteria. According to the valuation
protocol, interviews include practice and reference tasks that are not
intended for valuation analysis. These observations are therefore
excluded from the data set using the variable \texttt{state\_id}. To
enable regression-based valuation modelling, dummy variables are
generated for the five EQ-5D-5L dimensions (mo, sc, ua, pd, ad). Level 1
(``no problems'') serves as the reference category, resulting in dummy
indicators for levels 2--5 in each dimension. The resulting data set
comprises 17,336 observations from a total of 1,031 participants and
includes 36 variables. Each participant contributed seven DCE
observations and between seven and ten TTO observations.

\FloatBarrier

\begin{CodeChunk}
\begin{CodeInput}
R> library(EQ5Ddata)
R> library(fastDummies)
R> 
R> 
R> TTOonly <- hyregdata[hyregdata$method == "TTO" & 
+                        hyregdata$fb_flagged == 0 &
+                        hyregdata$state_id > 0,]
R> DCEonly <- hyregdata[hyregdata$method == "DCE_A" & 
+                        hyregdata$state_id < 197,]
R> data <- rbind(TTOonly,DCEonly)
R> 
R> 
R> data <- dummy_cols(
+   data,
+   select_columns = c("mo", "sc", "ua", "pd", "ad"),
+   remove_first_dummy = TRUE,   
+   remove_selected_columns = FALSE
+   )
R> 
R> colnames(data) <- gsub("_", "", colnames(data))
R> data <- data[,c(1:15,59:78,58)]
R> 
R> head(data)
\end{CodeInput}
\begin{CodeOutput}
    id VAS blockid stateid orderid mo sc ua pd ad originalresponse value method
1 2560  80       9      66       5  2  5  3  3  1               16   0.4    TTO
2 2560  80       9      67      13  2  5  2  2  2               14   0.6    TTO
3 2560  80       9      68       9  2  1  4  4  4               12   0.8    TTO
4 2560  80       9      69      14  3  1  5  1  4               13   0.7    TTO
5 2560  80       9      70      11  5  3  2  4  3               13   0.7    TTO
6 2560  80       9      71       8  5  3  2  4  4               11   0.9    TTO
  profile severity mo2 mo3 mo4 mo5 sc2 sc3 sc4 sc5 ua2 ua3 ua4 ua5 pd2 pd3 pd4
1   25331       14   1   0   0   0   0   0   0   1   0   1   0   0   0   1   0
2   25222       13   1   0   0   0   0   0   0   1   1   0   0   0   1   0   0
3   21444       15   1   0   0   0   0   0   0   0   0   0   1   0   0   0   1
4   31514       14   0   1   0   0   0   0   0   0   0   0   0   1   0   0   0
5   53243       17   0   0   0   1   0   1   0   0   1   0   0   0   0   0   1
6   53244       18   0   0   0   1   0   1   0   0   1   0   0   0   0   0   1
  pd5 ad2 ad3 ad4 ad5 dmethod
1   0   0   0   0   0       1
2   0   1   0   0   0       1
3   0   0   0   1   0       1
4   0   0   0   1   0       1
5   0   0   1   0   0       1
6   0   0   0   1   0       1
\end{CodeOutput}
\end{CodeChunk}

\subsection{Comparing hyreg2 and xreg in the one component
case}\label{comparing-hyreg2-and-xreg-in-the-one-component-case}

Before we can start to estimate the first model, the package has to be
installed and loaded using \texttt{install.packages("hyreg2")} and
\texttt{library(hyreg2)}. We use \texttt{k\ =\ 1} and compare the
results from our\pkg{hyreg2} package with those from the \pkg{xreg}
package. We use the regression model as given in equation \ref{eq.eq5d}.
The argument \texttt{formula} specifies the regression model, while
\texttt{data} contains the data set used for estimation. The input
parameter \texttt{type} expects either a column name of the data set as
\texttt{charachter} or a whole \texttt{vector}. \texttt{Type} identifies
the observation type and must distinguish between continuous and
dichotomous data. These are defined via \texttt{type\_cont} and
\texttt{type\_dich}. The argument \texttt{k} specifies the number of
latent classes. The optimization method is defined by
\texttt{opt\_method}, and additional optimization settings can be passed
via \texttt{control}.The input argument \texttt{latent} determines on
which type of data latent classes are expected to occur. Start values
must be given as a named vector \texttt{stv}. For this vector it is
essential that values are provided for all relevant independent
variables from the model \texttt{formula} including a possible intercept
or interaction terms. Using
\texttt{colnames(model.matrix(formula,data)),} one can determine for
which variables start values are expected. Additionally, start values
should be given for \texttt{sigma} and \texttt{theta} and these must be
named \texttt{"sigma"} and \texttt{"theta"}. If these are missing, they
are set to 1 by the function.

\FloatBarrier

\begin{CodeChunk}
\begin{CodeInput}
R> library(hyreg2)
R> 
R> 
R> formula <- value ~  mo2 + sc2 + ua2 + pd2 + ad2 + mo3 + sc3 + ua3 + pd3 + ad3 +
+                     mo4 + sc4 + ua4 + pd4 + ad4 + mo5 + sc5 + ua5 + pd5 + ad5
R> 
R> stvint <- setNames(c(rep(0.1,20),1,1,1),
+                    c(colnames(data)[16:35],c("sigma","theta","(Intercept)"))
+                    )
R> 
R> set.seed(25937)
R> mod1 <- hyreg2(formula = formula,
+                data = data,
+                type = data$method,
+                stv = stvint,
+                k = 1,
+                type_cont = "TTO",
+                type_dich = "DCE_A",
+                opt_method = "L-BFGS-B",
+                control = list(iter.max = 10000, verbose = 0),
+                latent = "both"
+                )
\end{CodeInput}
\end{CodeChunk}

\begin{CodeChunk}
\begin{CodeInput}
R> summary_hyreg2(mod1)
\end{CodeInput}
\begin{CodeOutput}
$Comp.1
Maximum likelihood estimation

Call:
bbmle::mle2(minuslogl = logLik2, start = stv_new, method = opt_method, 
    optimizer = optimizer, lower = -Inf, upper = Inf)

Coefficients:
              Estimate Std. Error  z value     Pr(z)    
(Intercept)  0.0879735  0.0183924   4.7831 1.726e-06 ***
mo2          0.0355943  0.0175130   2.0325 0.0421078 *  
sc2          0.0314392  0.0169627   1.8534 0.0638205 .  
ua2          0.0549314  0.0178798   3.0723 0.0021245 ** 
pd2          0.0286209  0.0161054   1.7771 0.0755506 .  
ad2          0.0374976  0.0178085   2.1056 0.0352384 *  
mo3          0.0858034  0.0183758   4.6694 3.021e-06 ***
sc3          0.0667237  0.0194382   3.4326 0.0005978 ***
ua3          0.0666901  0.0186451   3.5768 0.0003478 ***
pd3          0.1031112  0.0196726   5.2414 1.594e-07 ***
ad3          0.1348365  0.0200108   6.7382 1.604e-11 ***
mo4          0.1877863  0.0196936   9.5354 < 2.2e-16 ***
sc4          0.1695927  0.0187126   9.0630 < 2.2e-16 ***
ua4          0.1791813  0.0191893   9.3376 < 2.2e-16 ***
pd4          0.2569105  0.0169829  15.1276 < 2.2e-16 ***
ad4          0.2408934  0.0186698  12.9028 < 2.2e-16 ***
mo5          0.2698757  0.0181992  14.8290 < 2.2e-16 ***
sc5          0.1977600  0.0179463  11.0196 < 2.2e-16 ***
ua5          0.1846299  0.0174098  10.6050 < 2.2e-16 ***
pd5          0.2972957  0.0187861  15.8253 < 2.2e-16 ***
ad5          0.2830974  0.0177496  15.9495 < 2.2e-16 ***
sigma       -0.6748671  0.0070321 -95.9697 < 2.2e-16 ***
theta       -1.3371241  0.1363972  -9.8032 < 2.2e-16 ***
---
Signif. codes:  0 '***' 0.001 '**' 0.01 '*' 0.05 '.' 0.1 ' ' 1

-2 log L: 25004.89 
\end{CodeOutput}
\end{CodeChunk}

The model produces parameter values, which can now be compared with
those obtained from the \texttt{xreg} package. Figure
\ref{fig:comparison} visualizes the model estimates by \pkg{hyreg2}
compared to those generated using \pkg{xreg}. The values match up to the
third decimal place. Subsequent deviations result from differences in
rounding across optimization steps during maximum likelihood estimation
and can be disregarded. This confirms that the \texttt{hyreg2}
implementation of the framework by \citet{RamosGoni.2017} is compatible
with the \pkg{xreg} implementation, which has already been used in
several studies.

\FloatBarrier

\begin{CodeChunk}
\begin{figure}

{\centering \includegraphics{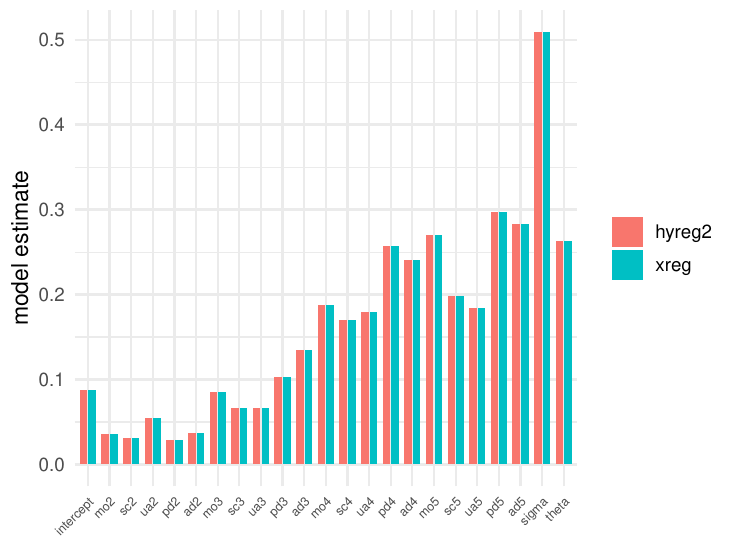} 

}

\caption[Comparison of estimates by \pkg{hyreg2} and \pkg{xreg}]{Comparison of estimates by \pkg{hyreg2} and \pkg{xreg}}\label{fig:comparison}
\end{figure}
\end{CodeChunk}

\subsection{Model estimation assuming two latent
classes}\label{model-estimation-assuming-two-latent-classes}

We now estimate two latent classes (\texttt{k\ =\ 2}) on the EQ-5D data
to further illustrate the additional functionality of \pkg{hyreg2}
compared to the \pkg{xreg} package and discuss the results. As explained
above and shown in the vignette, it is essential to control for groups
of observations. The column \texttt{id} contains the id numbers by
participants in the study and therefore indicates which observations
belong together. Accordingly, we control for these groups by using
\texttt{\textbar{}\ id} in the model \texttt{formula}. Moreover, it is
useful to specify \texttt{latent\ =\ “cont”}, since in this data set it
is assumed that latent classes are based only on the continuous data.
When using \texttt{latent\ =\ "cont",} the input parameter
\texttt{id\_col} must also be specified.

\FloatBarrier

\begin{CodeChunk}
\begin{CodeInput}
R> formula <- value ~ mo2 + sc2 + ua2 + pd2 + ad2 + mo3 + sc3 + ua3 + pd3 + ad3 +
+                    mo4 + sc4 + ua4 + pd4 + ad4 + mo5 + sc5 + ua5 + pd5 + ad5 | id
R> 
R> stv <- setNames(c(rep(0.1,20),1,1,1),
+                 c(colnames(data)[16:35],c("theta","sigma","(Intercept)"))
+                 )
R> 
R> set.seed(25937)
R> mod2 <- hyreg2(formula = formula,
+                data = data,
+                type = data$method,
+                stv = stv,
+                k = 2,
+                type_cont = "TTO",
+                type_dich = "DCE_A",
+                opt_method = "L-BFGS-B",
+                control = list(iter.max = 10000, verbose = 0),
+                latent = "cont",
+                id_col = "id"
+                )
\end{CodeInput}
\end{CodeChunk}

As output, the function now returns a \texttt{list} consisting of
\texttt{k\ +\ 1} elements: \texttt{k} fitted models and one data frame
containing two columns with the assigned class memberships for each
observation group (\texttt{id\_col}). The functions
\texttt{summary\_hyreg2} and \texttt{plot\_hyreg2} automatically detect
whether the model output is a single object or a list; therefore, no
adjustments to the function inputs are required.

We examine the obtained values in relation to the severity of the
overall underlying health state (i.e., across all dimensions). The
severity of a health state is defined as the sum of the levels across
all dimensions \citep{Devlin.2022}. For example, the health state 12221
has a severity of 8 (since 1 + 2 + 2 + 2 + 1 = 8). For this purpose, we
create a variable severity in the data set and subsequently relate the
model-based classification results to the corresponding severity levels.
To do so, we use \texttt{plot\_hyreg2}, which is described in more
detail below.

The function \texttt{plot\_hyreg2} is based on \pkg{ggplot2}
\citep{ggplot2} and returns a ggplot object, which can be further
customized using standard \texttt{ggplot} arguments and extensions. The
function requires the original data set (\texttt{data}), the variable to
be displayed on the x-axis as character (\texttt{x\ =\ "severity"}), and
the outcome variable for the y-axis as character
(\texttt{y\ =\ "value"}). The argument \texttt{id\_col} specifies the
grouping variable used in the model estimation and expects a character
\texttt{(id\_col\ =\ "id"}). The input parameter
\texttt{class\_df\_model} specifies the class assignment of the data
points. It expects a data frame with two columns. The first column
contains the groups of the grouping variable defined in
\texttt{id\_col}, while the second column indicates the class assigned
to each group by the model. This data frame can be created manually.
Alternatively, the function \texttt{give\_class()} can be used to
generate such a data frame. To restrict the visualization to a specific
type of data, we can use the \texttt{type\_to\_plot} argument. This
expects a list of two character elements. The first element must be a
column name from the data set, and the second element must be the column
value whose associated data should be visualized. Here we use
\texttt{list("method",\ "TTO")} and thereby we only plot observations
corresponding to the TTO data (i.e., observations with an entry of
``TTO'' in the \texttt{method} column). Using the \texttt{colors}
argument, custom class-specific colors can be defined.

\FloatBarrier

\begin{CodeChunk}
\begin{CodeInput}
R> data$severity  <- data$mo + data$sc + data$ua + data$pd + data$ad
R> 
R> plot_hyreg2(data = data,
+            x = "severity",
+            y = "value",
+            id_col = "id",
+            class_df_model = give_class(data = data,
+                                  model = mod2,
+                                  id_col = "id"),
+            type_to_plot = list("method","TTO"),
+            colors = c("#00BFC4","#F8766D")
+            )
\end{CodeInput}
\begin{figure}

{\centering \includegraphics{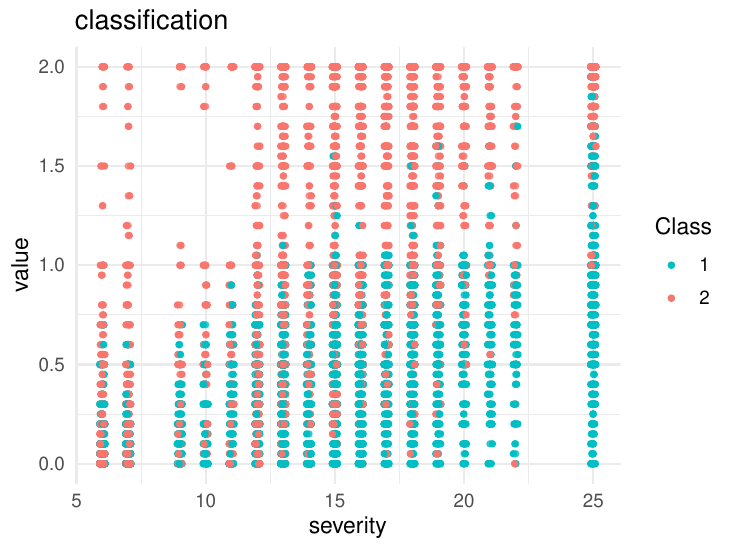} 

}

\caption[Classification according to severity level of the assessed health states]{Classification according to severity level of the assessed health states}\label{fig:plot}
\end{figure}
\end{CodeChunk}

Figure \ref{fig:plot} shows a clear pattern: participants in class 2
tend to reject health states (i.e.~score values between 1 and 2) more
frequently and at lower levels of impairment compared to those in class
1. This is reflected in the higher values on the y-axis. Within the
plot, an imaginary diagonal can be assumed, which appears to separate
the two classes. This demonstrates that the \texttt{hyreg2} package is
able to account for preference heterogeneity in the continuous data
part, as it partitions the data into two classes that clearly differ in
how people evaluate potential health impairments.

\subsection{Using censored data}\label{using-censored-data}

To account for censored data, the input parameters \texttt{upper} and/or
\texttt{lower} can be specified. These indicate the values at which the
entered dependent variable is censored. By default,
\texttt{upper\ =\ Inf} and \texttt{lower\ =\ -Inf}. To estimate models
with censored data, \texttt{opt\_method} must be set to
\texttt{"L-BFGS-B"}; otherwise, an error will occur. All other input
parameters can be used without modification.

\FloatBarrier

\begin{CodeChunk}
\begin{CodeInput}
R> formula <- value ~  mo2 + sc2 + ua2 + pd2 + ad2 + mo3 + sc3 + ua3 + pd3 + ad3 +
+                     mo4 + sc4 + ua4 + pd4 + ad4 + mo5 + sc5 + ua5 + pd5 + ad5 | id
R> 
R> stvint <- setNames(c(rep(0.1,20),1,1,1),
+                    c(colnames(data)[16:35],c("sigma","theta","(Intercept)"))
+                    )
R> 
R> set.seed(25937)
R> mod_cens <- hyreg2(formula = formula,
+                data = data,
+                type = data$method,
+                stv = stvint,
+                k = 2,
+                type_cont = "TTO",
+                type_dich = "DCE_A",
+                opt_method = "L-BFGS-B",
+                control = list(iter.max = 10000, verbose = 0),
+                upper = 2,
+                lower = 0,
+                latent = "cont",
+                id_col = "id"
+                )
\end{CodeInput}
\end{CodeChunk}

\subsection{Accounting for
heteroscedasticity}\label{accounting-for-heteroscedasticity}

Using this EQ-5D data set, we now demonstrate how \texttt{hyreg2\_het}
can be used to account for heteroscedasticity in the continuous data.
\texttt{hyreg2\_het} is structured analogously to \texttt{hyreg2}, but
includes two additional input parameters: \texttt{formula\_sigma} and
\texttt{stv\_sigma}. These can be used to explicitly model and estimate
the variance of the continuous data. \texttt{formula\_sigma} must be a
linear formula, but it may contain different predictors than
\texttt{formula}. For all variables included in \texttt{formula\_sigma},
starting values must be provided in \texttt{stv\_sigma} (analogous to
the specification of \texttt{formula} and \texttt{stv} described above).
Importantly, when using \texttt{hyreg2\_het}, no starting value for
\texttt{sigma} can be specified in \texttt{stv}. Since the variance
\texttt{sigma} is modeled explicitly in this case, a separate starting
value for \texttt{sigma} is not allowed.

\FloatBarrier

\begin{CodeChunk}
\begin{CodeInput}
R> formula <- value ~ mo2 + sc2 + ua2 + pd2 + ad2 + mo3 + sc3 + ua3 + pd3 + ad3 +
+                    mo4 + sc4 + ua4 + pd4 + ad4 + mo5 + sc5 + ua5 + pd5 + ad5 | id
R> 
R> stv <- setNames(c(rep(0.1,20),1,1),
+                 c(colnames(data)[16:35],c("theta","(Intercept)"))
+                 )
R> 
R> formula_sigma <- value ~  mo2 + sc2 + ua2 + pd2 + ad2 +
+                           mo3 + sc3 + ua3 + pd3 + ad3 +
+                           mo4 + sc4 + ua4 + pd4 + ad4 + 
+                           mo5 + sc5 + ua5 + pd5 + ad5
R> 
R> stv_sigma <- setNames(c(rep(0.1,20),1),
+                       c(colnames(data)[16:35],c("(Intercept)"))
+                       )
R> 
R> set.seed(25937)
R> mod_het <- hyreg2_het(formula = formula,
+                    formula_sigma = formula_sigma,
+                    data = data,
+                    type = data$method,
+                    stv = stv,
+                    stv_sigma = stv_sigma,
+                    k = 1,
+                    type_cont = "TTO",
+                    type_dich = "DCE_A",
+                    opt_method = "L-BFGS-B",
+                    control = list(iter.max = 10000, verbose = 0),
+                    latent = "cont",
+                    id_col = "id"
+                    )
\end{CodeInput}
\end{CodeChunk}

\begin{CodeChunk}
\begin{CodeInput}
R> summary_hyreg2(mod_het)
\end{CodeInput}
\begin{CodeOutput}
[[1]]
[[1]]$Comp.1
Maximum likelihood estimation

Call:
bbmle::mle2(minuslogl = logLik2, start = stv_new, method = opt_method, 
    optimizer = optimizer, lower = lower, upper = upper)

Coefficients:
               Estimate Std. Error  z value     Pr(z)    
(Intercept)    0.074205   0.012937   5.7357 9.709e-09 ***
mo2            0.040244   0.013828   2.9103 0.0036113 ** 
sc2            0.045251   0.013217   3.4238 0.0006175 ***
ua2            0.049886   0.013847   3.6026 0.0003151 ***
pd2            0.035389   0.012006   2.9475 0.0032036 ** 
ad2            0.038447   0.012488   3.0786 0.0020795 ** 
mo3            0.084942   0.018061   4.7031 2.563e-06 ***
sc3            0.077494   0.016799   4.6129 3.971e-06 ***
ua3            0.066775   0.016746   3.9874 6.680e-05 ***
pd3            0.098976   0.018791   5.2672 1.385e-07 ***
ad3            0.128113   0.018421   6.9548 3.531e-12 ***
mo4            0.183416   0.018643   9.8385 < 2.2e-16 ***
sc4            0.184425   0.018230  10.1166 < 2.2e-16 ***
ua4            0.187502   0.018004  10.4145 < 2.2e-16 ***
pd4            0.260911   0.016767  15.5608 < 2.2e-16 ***
ad4            0.250794   0.017370  14.4380 < 2.2e-16 ***
mo5            0.265786   0.017118  15.5267 < 2.2e-16 ***
sc5            0.220467   0.016338  13.4938 < 2.2e-16 ***
ua5            0.184719   0.016859  10.9569 < 2.2e-16 ***
pd5            0.312880   0.018268  17.1274 < 2.2e-16 ***
ad5            0.287479   0.015743  18.2604 < 2.2e-16 ***
theta         -1.326370   0.132649  -9.9991 < 2.2e-16 ***
mo2_h          0.092974   0.024237   3.8361 0.0001250 ***
mo3_h          0.192395   0.025017   7.6906 1.464e-14 ***
mo4_h          0.203967   0.027528   7.4096 1.267e-13 ***
mo5_h          0.215527   0.023741   9.0782 < 2.2e-16 ***
sc2_h          0.082375   0.023757   3.4674 0.0005256 ***
sc3_h          0.155525   0.027310   5.6949 1.234e-08 ***
sc4_h          0.209492   0.025078   8.3535 < 2.2e-16 ***
sc5_h          0.170768   0.023505   7.2653 3.722e-13 ***
ua2_h          0.108936   0.026008   4.1886 2.807e-05 ***
ua3_h          0.149901   0.026428   5.6722 1.410e-08 ***
ua4_h          0.241497   0.026803   9.0101 < 2.2e-16 ***
ua5_h          0.133620   0.023278   5.7401 9.462e-09 ***
pd2_h          0.065361   0.022626   2.8888 0.0038672 ** 
pd3_h          0.167472   0.027427   6.1061 1.021e-09 ***
pd4_h          0.187242   0.023083   8.1117 4.992e-16 ***
pd5_h          0.197602   0.024939   7.9234 2.311e-15 ***
ad2_h          0.067702   0.024792   2.7308 0.0063180 ** 
ad3_h          0.193212   0.027212   7.1003 1.245e-12 ***
ad4_h          0.287222   0.025695  11.1780 < 2.2e-16 ***
ad5_h          0.213760   0.023442   9.1188 < 2.2e-16 ***
(Intercept)_h -1.352016   0.026986 -50.1015 < 2.2e-16 ***
---
Signif. codes:  0 '***' 0.001 '**' 0.01 '*' 0.05 '.' 0.1 ' ' 1

-2 log L: 23879.66 
\end{CodeOutput}
\end{CodeChunk}

The summary output now provides not only the estimated model parameters
(from \texttt{formula}) but also the parameters for estimating the
varaince \texttt{sigma}. These can be identified by the ending
``\texttt{\_h}''. The \pkg{xreg} package also offers the possibility to
account for heteroscedasticity. Again, the same results are provided for
\texttt{hyreg2\_het} and \texttt{xreg} using \texttt{k\ =\ 1}.

\subsection{Using non-classic model
formulas}\label{using-non-classic-model-formulas}

All inputs and functionalities that have been considered so far utilize
classic R formulas and the default setting (\texttt{TRUE}) for the input
\texttt{formula\_type\_classic}. However, with this setting, it is not
possible to estimate non-linear formulas (e.g.~equation \ref{eq:nonlin})
or the so called 8-parameter model for EQ-5D data as given in equation
\ref{eq:eq8p} \citep{Devlin.2022, Luo.2017}.

\vspace{1.5em}

\begin{equation}
value = \frac{1}{e^{\beta_{0} + mo_2 * \beta_{mo2} +mo_3 *  \beta_{mo3}}} 
\label{eq:nonlin}
\end{equation} \begin{equation}
\begin{aligned}
value &= (mo2 * \beta_{mo} + sc2 * \beta_{sc} + ua2 * \beta_{ua} + pd2 * \beta_{pd} + ad2 * \beta_{ad}) * \beta_{L2} \\
&+ (mo3 * \beta_{mo} + sc3 * \beta_{sc} + ua3 * \beta_{ua} + pd3 * \beta_{pd} + ad3 * \beta_{ad}) * \beta_{L3} \\
&+ (mo4 * \beta_{mo} + sc4 * \beta_{sc} + ua4 * \beta_{ua} + pd4 * \beta_{pd} + ad4 * \beta_{ad}) * \beta_{L4} \\
&+ (mo5 * \beta_{mo} + sc5 * \beta_{sc} + ua5 * \beta_{ua} + pd5 * \beta_{pd} + ad5 * \beta_{ad}) \\
\end{aligned}
\label{eq:eq8p}
\end{equation}

\vspace{1.5em}

Therefore, in \texttt{hyreg2}, we explicitly included the option of
specifying model formulas with variables and parameters (non-classic
model formulas). Please note that, currently, this functionality is only
available for \texttt{hyreg2} and not for \texttt{hyreg2\_het}. Table
\ref{tab:nonclassic} compares classic and non-classic model formulas to
highlight the conceptual differences between the two approaches. In the
classic formula specification, the column names of the data set
correspond directly to the parameters to be estimated, meaning that
model coefficients are associated with the column names of the data set.
In contrast, non-classic formulas follow a representation that is closer
to standard mathematical notation. In this case, parameters (in this
case called \texttt{beta}) must be specified explicitly in addition to
the data sets column names. These parameters are not themselves
variables in the data set but represent model coefficients that are
estimated during the model estimation process. For classical linear
regression models, such as the 20-parameter model specified in equation
\ref{eq.eq5d}, both classic and non-classic formula specifications can
be used in \pkg{hyreg2}, although the classic specification is generally
preferable. In contrast, for models that want to estimate the same
parameter using multiple columns of the data set (e.g.~parameter
\(\beta_{mo}\) in the 8-parameter model given in equation
\ref{eq:eq8p}), \pkg{hyreg2} requires the use of non-classic formulas.
The same applies to non-linear models, such as the model specified in
equation \ref{eq:nonlin}, for which only non-classic formula will work.

\begin{table}[htbp]
\centering
\small
\renewcommand{\arraystretch}{1.6}
\setlength{\extrarowheight}{4pt}

\begin{tabular}{>{\raggedright\arraybackslash}p{3.5cm}
                >{\raggedright\arraybackslash}p{5.5cm}
                >{\raggedright\arraybackslash}p{7cm}}
\toprule
Equation & Classic model formula & Non-classic model formula \\
\midrule

20-parameter model, equation \ref{eq.eq5d} &
\texttt{value ~ mo2 + sc2 + ua2 + pd2 + ad2 + mo3 + sc3 + ua3 + pd3 + ad3 + mo4 + sc4 + ua4 + pd4 + ad4 + mo5 + sc5 + ua5 + pd5 + ad5}
&
\texttt{value ~ mo2 * beta\_mo2 + sc2 * beta\_sc2 + ua2 * beta\_ua2 + pd2 * beta\_pd2 + ad2 * beta\_ad2 + mo3 * beta\_mo3 + sc3 * beta\_sc3 + ua3 * beta\_ua3 + pd3 * beta\_pd3 + ad3 * beta\_ad3 + mo4 * beta\_mo4 + sc4 * beta\_sc4 + ua4 * beta\_ua4 + pd4 * beta\_pd4 + ad4 * beta\_ad4 + mo5 * beta\_mo5 + sc5 * beta\_sc5 + ua5 * beta\_ua5 + pd5 * beta\_pd5 + ad5 * beta\_ad5} \\

8-parameter model, equation \ref{eq:eq8p} &
- &
\texttt{value ~ (mo2 * beta\_mo + sc2 * beta\_sc + ua2 * beta\_ua + pd2 * beta\_pd + ad2 * beta\_ad) * beta\_L2 + (mo3 * beta\_mo + sc3 * beta\_sc + ua3 * beta\_ua + pd3 * beta\_pd + ad3 * beta\_ad) * beta\_L3 + (mo4 * beta\_mo + sc4 * beta\_sc + ua4 * beta\_ua + pd4 * beta\_pd + ad4 * beta\_ad) * beta\_L4 + (mo5 * beta\_mo + sc5 * beta\_sc + ua5 * beta\_ua + pd5 * beta\_pd + ad5 * beta\_ad)} \\

non-linear model, equation \ref{eq:nonlin} &
- & 
\texttt{value ~ 1/exp(INTERCEPT + mo2 * beta\_mo2 + mo3 * beta\_mo3)} \\

\bottomrule
\end{tabular}
\caption{Comparison of classic and non-classic model formulas.}
\label{tab:nonclassic}
\end{table}

When estimating models using non-classic model formulas,
\texttt{formula\_type\_classic} must be set to \texttt{FALSE}. When
estimating an intercept, the formula must explicitly include a parameter
named \texttt{"INTERCEPT"}(without a corresponding variable from the
data set),
e.g.~\texttt{formula\ =\ value\ \textasciitilde{}\ \ 1/exp(INTERCEPT\ +\ mo2\ *\ beta\_mo2\ +\ mo3\ *\ beta\_mo3)}.
The starting values \texttt{stv} must be provided as a named vector. The
names of this vector must correspond to the parameters to be estimated
(i.e., the \texttt{beta} coefficients) and not to the column names of
the data set.

\FloatBarrier

\begin{CodeChunk}
\begin{CodeInput}
R> formula_nonc <- value ~ (mo2 * beta_mo + sc2 * beta_sc  + ua2 * beta_ua +
+                            pd2 * beta_pd + ad2 * beta_ad) *   beta_L2 +
+                         (mo3 * beta_mo + sc3 * beta_sc  + ua3 * beta_ua + 
+                            pd3 * beta_pd + ad3 * beta_ad) * beta_L3 +
+                         (mo4 * beta_mo + sc4 * beta_sc  + ua4 * beta_ua +
+                            pd4 * beta_pd + ad4 * beta_ad) * beta_L4 +
+                         (mo5 * beta_mo + sc5 * beta_sc  + ua5 * beta_ua +
+                            pd5 * beta_pd + ad5 * beta_ad) | id
R> 
R> stv_nonc <- setNames(c(rep(0.1,8),1,1),
+                      c("beta_mo","beta_sc", "beta_ua","beta_pd","beta_ad",
+                        "beta_L2", "beta_L3","beta_L4","sigma","theta")
+                      )
R> 
R> set.seed(25937)
R> mod_nonc <- hyreg2(formula = formula_nonc,
+                data = data,
+                type = data$method,
+                stv = stv_nonc,
+                k = 2,
+                type_cont = "TTO",
+                type_dich = "DCE_A",
+                opt_method = "L-BFGS-B",
+                control = list(iter.max = 10000, verbose = 0),
+                latent = "cont",
+                id_col = "id",
+                formula_type_classic = FALSE
+                )
\end{CodeInput}
\end{CodeChunk}

\begin{CodeChunk}
\begin{CodeInput}
R> summary_hyreg2(mod_nonc)
\end{CodeInput}
\begin{CodeOutput}
[[1]]
[[1]]$Comp.1
Maximum likelihood estimation

Call:
bbmle::mle2(minuslogl = logLik2, start = stv_new, method = opt_method, 
    optimizer = optimizer, lower = -Inf, upper = Inf)

Coefficients:
          Estimate Std. Error   z value     Pr(z)    
beta_mo  0.1705509  0.0105907   16.1038 < 2.2e-16 ***
beta_sc  0.1558334  0.0107825   14.4525 < 2.2e-16 ***
beta_ua  0.1650798  0.0102358   16.1277 < 2.2e-16 ***
beta_pd  0.2208641  0.0107151   20.6124 < 2.2e-16 ***
beta_ad  0.2176210  0.0103328   21.0613 < 2.2e-16 ***
beta_L2  0.2475729  0.0179466   13.7950 < 2.2e-16 ***
beta_L3  0.3416177  0.0226128   15.1073 < 2.2e-16 ***
beta_L4  0.7004480  0.0202861   34.5285 < 2.2e-16 ***
sigma   -1.3232061  0.0091173 -145.1321 < 2.2e-16 ***
theta   -0.9096106  0.1763245   -5.1587 2.486e-07 ***
---
Signif. codes:  0 '***' 0.001 '**' 0.01 '*' 0.05 '.' 0.1 ' ' 1

-2 log L: 7038.068 


[[2]]
[[2]]$Comp.1
Maximum likelihood estimation

Call:
bbmle::mle2(minuslogl = logLik2, start = stv_new, method = opt_method, 
    optimizer = optimizer, lower = -Inf, upper = Inf)

Coefficients:
         Estimate Std. Error  z value     Pr(z)    
beta_mo  0.410793   0.024128  17.0256 < 2.2e-16 ***
beta_sc  0.283320   0.025452  11.1314 < 2.2e-16 ***
beta_ua  0.310263   0.023265  13.3363 < 2.2e-16 ***
beta_pd  0.463993   0.024246  19.1371 < 2.2e-16 ***
beta_ad  0.408146   0.024533  16.6363 < 2.2e-16 ***
beta_L2  0.297658   0.021896  13.5940 < 2.2e-16 ***
beta_L3  0.532491   0.027941  19.0577 < 2.2e-16 ***
beta_L4  0.949000   0.026455  35.8725 < 2.2e-16 ***
sigma   -0.620142   0.011042 -56.1617 < 2.2e-16 ***
theta   -1.793736   0.211809  -8.4686 < 2.2e-16 ***
---
Signif. codes:  0 '***' 0.001 '**' 0.01 '*' 0.05 '.' 0.1 ' ' 1

-2 log L: 10605.97 
\end{CodeOutput}
\end{CodeChunk}

The model now returns estimated values for the specified parameters. In
contrast to classic formulas, the column names used in the model do not
appear in the summary output, as no separate coefficients are estimated
for them in non-classic model formulas.

\section{Conclusion and Outlook}\label{conclusion-and-outlook}

The R package \pkg{hyreg2} represents a meaningful extension of existing
software for the estimation of joint likelihood models, as it enables
user-friendly estimation of latent class models within this framework.
It provides accessible options to estimate more complex model
specifications, such as heteroscedastic or non-linear models. The
simulation study of the vignette and the case study presented in this
article demonstrate that \pkg{hyreg2} produces plausible and meaningful
results. Although the model proposed by \citet{RamosGoni.2017} was
originally developed for applications in health economics, it can also
be applied in many other scientific fields. This is particularly
relevant whenever different types of data need to be analysed together
within one model.

It should be noted that---as is common in maximum likelihood
estimation---the results may be highly sensitive to the choice of
starting values. Selecting appropriate initial values is therefore
particularly important. Whenever possible, we recommend choosing
starting values based on the existing literature. In addition, applying
a bootstrapping procedure is advisable to assess the robustness of the
results and to distinguish between local and global maxima.

The package is fully functional but still offers potential for further
development. At present, non-classical functions are supported only for
the \texttt{hyreg2} function, but not yet for \texttt{hyreg2\_het}.
Furthermore, random intercepts are currently not supported. A function
for automated model selection based on the \pkg{flexmix} framework would
also be useful. Finally, it would be possible to extend the package by
implementing additional joint likelihood specifications, allowing users
to select from a list of distributions that can be combined within the
likelihood function. Such an extension would broaden the potential user
base of the package, as it would move beyond the specific model proposed
by \citet{RamosGoni.2017}.

\section{Acknowledgements}\label{acknowledgements}

The development of the \pkg{hyreg2} package was supported by the EuroQol
Research Foundation under grant number 209-RA.

\section{Computational details}\label{computational-details}

\begin{CodeChunk}
\begin{CodeOutput}
R version 4.5.3 (2026-03-11 ucrt)
Platform: x86_64-w64-mingw32/x64
Running under: Windows 11 x64 (build 26100)

Matrix products: default
  LAPACK version 3.12.1

locale:
[1] LC_COLLATE=German_Germany.utf8  LC_CTYPE=German_Germany.utf8   
[3] LC_MONETARY=German_Germany.utf8 LC_NUMERIC=C                   
[5] LC_TIME=German_Germany.utf8    

time zone: Europe/Berlin
tzcode source: internal

attached base packages:
[1] stats     graphics  grDevices utils     datasets  methods   base     

other attached packages:
[1] ggplot2_4.0.2     xreg_0.3.1        hyreg2_1.1.2      fastDummies_1.7.5
[5] EQ5Ddata_0.1.0   

loaded via a namespace (and not attached):
 [1] Matrix_1.7-4        gtable_0.3.6        compiler_4.5.3     
 [4] stringr_1.6.0       scales_1.4.0        yaml_2.3.12        
 [7] fastmap_1.2.0       lattice_0.22-9      R6_2.6.1           
[10] labeling_0.4.3      bbmle_1.0.25.1      knitr_1.51         
[13] MASS_7.3-65         bdsmatrix_1.3-7     tibble_3.3.1       
[16] nnet_7.3-20         RColorBrewer_1.1-3  pillar_1.11.1      
[19] rlang_1.1.7         stringi_1.8.7       xfun_0.56          
[22] S7_0.2.1            modeltools_0.2-24   otel_0.2.0         
[25] cli_3.6.5           withr_3.0.2         magrittr_2.0.4     
[28] digest_0.6.39       grid_4.5.3          rstudioapi_0.18.0  
[31] mvtnorm_1.3-3       rticles_0.27        flexmix_2.3-20     
[34] lifecycle_1.0.5     vctrs_0.7.1         evaluate_1.0.5     
[37] glue_1.8.0          data.table_1.18.2.1 farver_2.1.2       
[40] numDeriv_2016.8-1.1 stats4_4.5.3        rmarkdown_2.30     
[43] tools_4.5.3         pkgconfig_2.0.3     htmltools_0.5.9    
\end{CodeOutput}
\end{CodeChunk}

\bibliographystyle{plainnat}
\bibliography{hyreg2_arxiv}

\end{document}